\documentclass[aps,pra,amssymb,onecolumn,showpacs,a4paper]{revtex4-1}

\usepackage{amsmath,amssymb,epsfig,graphicx,color,framed}
\usepackage{hyperref}
\usepackage{amsfonts} 
\usepackage{amsmath}
\usepackage{amssymb}
\usepackage{graphicx}
\usepackage{hyperref}
\usepackage{color}
\usepackage{natbib}  
\usepackage{bm} 
\usepackage{isomath} 
\usepackage{braket}
\usepackage{booktabs} 

\begin{document}

\title{Time-delayed Duffing oscillator in an active bath} 

\author{Antonio A. Valido}
\email{alejandro.valido@urjc.es}
\affiliation{Nonlinear Dynamics, Chaos and Complex Systems Group, Departamento de F\'{i}sica,
Universidad Rey Juan Carlos, Tulip\'{a}n s/n, 28933 M\'{o}stoles, Madrid, Spain}

\author{Mattia Coccolo}
\affiliation{Nonlinear Dynamics, Chaos and Complex Systems Group, Departamento de F\'{i}sica,
Universidad Rey Juan Carlos, Tulip\'{a}n s/n, 28933 M\'{o}stoles, Madrid, Spain}

\author{Miguel A.F. Sanju\'{a}n}
\affiliation{Nonlinear Dynamics, Chaos and Complex Systems Group, Departamento de F\'{i}sica,
Universidad Rey Juan Carlos, Tulip\'{a}n s/n, 28933 M\'{o}stoles, Madrid, Spain}

\date{\today}

\begin{abstract}
During the last decades active particles have attracted an incipient attention as they have been observed in a broad class of scenarios, ranging from bacterial suspension in living systems to artificial swimmers in nonequilibirum systems. The main feature of these particles is that they are able to gain kinetic energy from the environment, which is widely modeled by a stochastic process due to both (Gaussian) white and Ornstein-Uhlenbeck noises. In the present work, we study the nonlinear dynamics of the forced, time-delayed Duffing oscillator subject to these noises, paying special attention to their impact upon the maximum oscillations amplitude and characteristic frequency of the steady state for different values of the time delay and the driving force. Overall, our results indicate that the role of the time delay is substantially modified with respect to the situation without noise.  For instance, we show that the oscillations amplitude grows with increasing noise strength when the time delay acts as a damping term in absence of noise, whereas the trajectories eventually become aperiodic when the oscillations are sustained by the time delay. In short, the interplay among the noises, forcing and time delay gives rise to a rich dynamics: a regular and periodic motion is destroyed or restored owing to the competition between the noise and the driving force depending on time delay values, whereas an erratic motion insensitive to the driving force emerges when the time delay makes the motion aperiodic. Interestingly, we also show that, for a sufficient noise strength and forcing amplitude, an approximately periodic interwell motion is promoted by means of stochastic resonance.

\end{abstract}

\maketitle

\section{Introduction}

During the last decades active matter have attracted considerable attention in a broad set of scientific disciplines, ranging from statistical physics to biology and chemical physics \cite{romanczuk20121,bechinger20161}. Basically, active matter consists of particles enabled to move autonomously by harnessing energy from the environment (or within the system) and converting it into directed motion or useful work \cite{ghosh20221}, such as swarming bacteria cells \cite{li20211,seyforth20221} or synthetic self-propelled colloids. These systems have been intensively investigated in the context of nonequilibrium statistical physics such as nonequilibrium fluctuations \cite{fodor20221} or many-body Langevin dynamics \cite{zottl20231}, among others.

Active systems can display remarkable phenomena such as collective behaviors, motility-induced phase separation, or equilibrium-like features \cite{bonilla20191,martin20211,byrne20221} (e.g., they fulfill a generalized equipartition relation under appropriate conditions \cite{maggi20141}) despite they are prevented from being at equilibrium with the environment (as the particle motion does not satisfy a fluctuation-dissipation relation owing to the intrinsic nonequilibrium character of activity). A prominent example of an active system is the so-called active Ornstein-Uhlenbeck (OU) particle \cite{crisanti20231,dabelow20211,nguyen20221,fodor20221}, that is characterized by an exponentially-correlated fluctuating force whose dynamics is described by an OU process, and is widely referred to as active or colored noise \cite{hanggi20071,kosek19881}. The latter has been used to model transport properties of active colloids \cite{nguyen20221} as well as collective cell dynamics \cite{deforet20141}, and has found numerous applications in different scientific fields beyond active systems \cite{kosek19881,hanggi20071}. Interestingly, it has been shown that active OU particles satisfy detailed balance in the presence of potentials with zero third derivatives \cite{bonilla20191,fodor20161}.

Despite the preceding theoretical and experimental progresses, nonlinear effects in active systems remains largely unexplored: most of the recent results are restricted to linear or quadratic potentials \cite{bechinger20161,kopp20231,szamel20141,ghosh20221}, inertial effects under a harmonic potential \cite{nguyen20221}, or transport properties in harmonic chains coupled to a reservoir composed of active OU \cite{sarkar20231} or run-and-tumble particles (coined as active reservoirs) \cite{santra20221}. It is expected that the introduction of nonlinearity can open new avenues of investigation \cite{zottl20231}: for instance, it has been recently analyzed the nonlinear rheology of active baths by experimental means \cite{seyforth20221}, as well as the escape rate of an anharmonic active particle \cite{sharma20171}. Moreover, it is well known that in realistic setups the forces and interactions involve some time lag (e.g. the time delay has been manifested as a sensorial delay in biological living systems \cite{glass20211}) that can be important in chaos control, vibrational resonance \cite{rajasekar20161,jeevarathinam20111} as well as delay-induced resonance \cite{cantisan20201,coccolo20211}. Although  strong similarities have been recently identified between the dynamics of a particle subject to a repulsive delayed feedback and active motion \cite{kopp20231}, the interplay between a time delay and an active noise has not yet been addressed to the best of our knowledge.

In this paper, we aim to contribute to fill this gap and explore the role of the time delay in active nonlinear particles. In particular, we study the impact of the active OU as well as white noises in the stationary dynamics of a forced, time-delayed Duffing oscillator; and provide an extensive numerical analysis of the maximum peak-to-peak oscillations amplitude response and the characteristic frequency of the oscillator. We find that the time-delayed nonlinear dynamics is similarly influenced by both noises, i.e., taken separately there is no significant distinction between the active and white noise effects in either the amplitude or the frequency. On the other hand, when the two noises are combined with an external forcing the effect is more complicated and worth to be analyzed. For instance, in contrast to the situation in absence of noise, we show that the limit cycle returned by certain values of the time delay gets destroyed in presence of both noises as the forcing amplitude grows. Conversely, the noise can effectively combine with the driving force to give rise to a regular and periodic dynamics despite the role played by certain time delay values that make it acts as an effective damping term.  Although our main purpose is to provide a fundamental analysis of the impact of perturbative noise effects upon the time-delayed dynamics, we also pay attention to the intense noise scenario. Interestingly, we show that, for a sufficient forcing amplitude, the oscillator may describe an approximately periodic interwell motion due to a synchronization between the noise action and the periodic forcing. This resembles the well-known stochastic resonance phenomenon previously exhibited by bistable systems subject to the white noise~\cite{gammaitoni19891,gammaitoni19981,choi19981,hanggi20001}. 

\section{Model and methods}\label{SecMMT}
The model under consideration consists of the one-dimensional dynamics of the forced, time-delayed Duffing oscillator subjected to both Gaussian white and OU noises \cite{romanczuk20121}. The stochastic equation that governs the underdamped dynamics for its spatial coordinate $x(t)$ reads 
\begin{equation}
    \frac{d^2 x(t)}{dt^2}+\mu\frac{dx(t)}{dt}+\frac{\partial U_{\tau}(x,t)}{\partial x}=F\cos(\Omega t)+\sqrt{2D_{wn}}\xi_{wn}(t)+\sqrt{2D_{ou}}\xi_{ou}(t),
    \label{EqDON}
\end{equation}
where $D_{wn}$, $D_{ou}$, $\mu$ and $F$ denote the strength of the white noise, the OU noise (also called colored noise), the friction coefficient, and the amplitude of the unbiased time-periodic force, respectively. Whereas $\Omega$ is the frequency of the time-periodic force and $U_{\tau}$ is the time-delayed double-well potential \cite{jeevarathinam20111}, i.e.
\begin{equation}
    U_{\tau}(x,t)=\beta\frac{x^4(t)}{4}-\frac{\alpha}{2}x^2(t)+\frac{\gamma}{2}x^2(t-\tau),
    \label{PtFunc}
\end{equation}
where the last term in the right hand represents the time-delayed feedback of strength $\gamma$ and time delay $\tau$, whilst $\alpha$ and $\beta$ denote the linear and nonlinear coupling coefficients. For sake of clarity, we shall consider a constant historical time delay function~\cite{cantisan20201} which cancels for $t<\tau$, i.e. $x(t-\tau) =0$ when $t<\tau$.  Besides the origin $x^{\ast}_{0}=0$, the double-well potential associated to the time-delayed Duffing oscillator has two other equilibrium points located at the bottom of the potential wells~\cite{rajasekar20161} 
\begin{equation}\label{eq:2}
x^*_{1,2}=\pm\sqrt{\frac{\alpha-\gamma}{\beta}},
\end{equation}
when $\alpha,\beta>0$ (otherwise $U_{\tau}$ may become a single well). These are separated by a potential barrier with the height determined by $\Delta U=(\alpha-\gamma)^2/(4\beta)$. In Refs.~\cite{cantisan20201,coccolo20211}  a linear stability analysis was performed taking into account the time-delayed term (by setting $x(t)=x(t-\tau)=x^\ast$) and ignoring the driving force around the points $x^{\ast}_{0}$, $x^{\ast}_{1}$, and $x^*_{2}$. This analysis reveals that their stability is substantially influenced by the values of the time delay $\tau$ and its strength $\gamma$  (see Sec.~\ref{SecPre} for further details). Here, we shall focus the attention on certain set of parameters for which $x^{\ast}_{0}$ remains unstable, whilst the stability of $x^*_{1,2}$ may change with $\tau$. For instance, this scenario has been employed to investigate both the stochastic~\cite{hanggi20001} and the vibrational resonances \cite{jeevarathinam20111,rajasekar20161,cantisan20201} (in the absence of noise and without time delay). The overdamped dynamics of an active particle subject to white and OU noises has been addressed in~\cite{bothe20211}.

In Eq.~(\ref{EqDON}), the stochastic zero-mean $\delta$-correlated Gaussian noise source $\xi_{wn}$, with $D_{wn}$ strength, represents the environmental fluctuations in the framework of active particles that is usually assumed in a thermal equilibrium state \cite{dabelow20211}. Different from the latter, $\xi_{ou}$ plays the role of the active self-propulsion \cite{maggi20141,dabelow20211,li20211,fodor20221}. This models a Gaussian-distributed random force exhibiting exponentially decaying memory effects characterized by a well-defined noise correlation time $\tau_{ou}$ \cite{martin20211,romanczuk20121,ghosh20221,nguyen20221}, and endowed with a time-asymptotic non-zero mean $\xi_{ou}^{\infty}$. Collecting these properties, these random variables are completely characterized by their mean values and covariance functions (that is, $\text{cov}\{\xi(t)\xi(t')\}=\langle(\xi(t)-\langle \xi(t)\rangle)(\xi(t')-\langle \xi(t')\rangle)\rangle$) when $0<t'<t$ \cite{gillespie19961}: i.e.,
\begin{align}
    \langle \xi_{wn}(t)\rangle_{wn}&=0 ,  \ \ \ \ \ \ \ \ \  \text{cov}\{\xi_{wn}(t)\xi_{wn}(t')\}_{wn}=C_{wn}\delta(t-t'), \label{EQWN} \\
    \langle \xi_{ou}(t)\rangle_{ou}&=\xi_{ou}^{0} e^{\frac{-t}{\tau_{ou}}} + \xi_{ou}^{\infty} \Big(1-e^{\frac{-t}{\tau_{ou}}}\Big) , \ \ \ \ \ \ \ \ \  \text{cov}\{\xi_{ou}(t)\xi_{ou}(t')\}_{ou}=\frac{C_{ou}\tau_{ou}}{2}e^{-\frac{(t-t')}{\tau_{ou}}}\Big(1- e^{-\frac{2t'}{\tau_{ou}}}\Big), \label{EQOUN}
\end{align}
with $\delta$ being the Dirac-delta function,  $\xi_{ou}^{0}$ stands for the initial mean value of the colored noise, whereas $C_{wn}$ and $C_{ou}$ are the amplitude of the white noise and the strength of the self-propulsion, respectively. In the limit of vanishing memory effects $\tau_{ou}\rightarrow 0$, $\xi_{ou}$ reduces to a Gaussian white noise. 

Over time scales that are larger than both the characteristic time scale of the periodic force, the system is expected to reach a steady state driven by both noise fluctuations together with the time delay potential feedback. By paying attention to the correlations functions in (\ref{EQWN}) and (\ref{EQOUN}), one may see that the white and colored noises have opposite effects along the oscillator trajectory: while the former represents a random disturbance washing out any memory effect, the later introduces noisy correlations enduring a time given by $\tau_{ou}$. Precisely, this memory effect describes the persistent motion exhibited by either passive tracer particles in an active bath \cite{maggi20171,maggi20141} or self-propelled particles \cite{nguyen20221}. For instance, though it is not considered here, if the environment is in a thermal-equilibrium state at certain temperature $T$, the oscillator would be expected to eventually reach a thermal equilibrium state since $\xi_{wn}$ would represent the standard thermal noise satisfying the fluctuation-dissipation theorem (i.e. $C_{wn}=k_{B}T\mu$ \cite{ghosh20221} with $k_{B}$ being the Boltzmann constant). However the oscillator is prevented from such equilibrium due to the (active) self-propulsion force modeled by $\xi_{ou}$ \cite{dabelow20211}. In other words, the physical effect is that the system itself is either generating the noise internally or it is subject to the noise as an external perturbation. 

Before to further proceed it is convenient to briefly describe the numerical simulation procedure. Concretely, we numerically solve the stochastic delay differential equation (\ref{EqDON}) for a sufficiently large time $T_{\infty}\ge 100\frac{2\pi}{\Omega}$ (which ensures the system has reached the stationary state) by using a method named S-ROCK described in Ref.~\cite{komori20191} and implemented in Matlab. This consists on a direct numerical integration of the equations of motion based on a Runge–Kutta-Chebyshev scheme of order four considering a fixed value of time grid $dt=10^{-3}$, which outperforms the previously employed Euler-Maruyama algorithms in Refs.~\cite{buckwar20001,du20111}. The white noise satisfying the standard Gaussian distribution are produced by the Box–Muller algorithm, whereas the colored noise is generated by means of the Gillespie algorithm \cite{gillespie19961} (which essentially consists of an algebraic manipulation and transformation on the white Gaussian random numbers). Notice that the angular brackets $\langle \bullet\rangle_{wn}$ and $\langle\bullet\rangle_{ou}$ indicate the average over these Gaussian probability distributions associated to the white and colored noises. In practice, this average is performed over an ensemble composed of a sufficiently large number $N_{sim}$ of trajectories with different noise realizations such that we recover Eq.~(\ref{EQWN}) and Eq.~(\ref{EQOUN}), but with identical initial conditions. Since Eq.~(\ref{EqDON}) has to be solved for the aforementioned long asymptotic time, this procedure can be computationally time consuming. For our simulation purposes, $N_{sim}=100$ turns to be sufficient to get reliable outcomes for the maximum peak-to-peak oscillations amplitude and characteristic frequency. More specifically, we have assessed the convergence of the algorithm for increasing $N_{sim}$, the results are reported in the supplementary figures found in Appendix~\ref{App1} (see Fig.~\ref{Fig_Tau_Noise_Conv}): basically, it is found that the algorithm is convergent and the numerical results hardly depend on the time step and the number of trajectories when $N_{sim}\gtrsim 100$. From now on, we set $\xi_{ou}^{0}=0$ as we are interested in the time asymptotic dynamics, and we shall use  dimensionless quantities.

\subsection{Preliminary results}\label{SecPre}
\begin{figure*}
\centering
\includegraphics[scale=0.55]{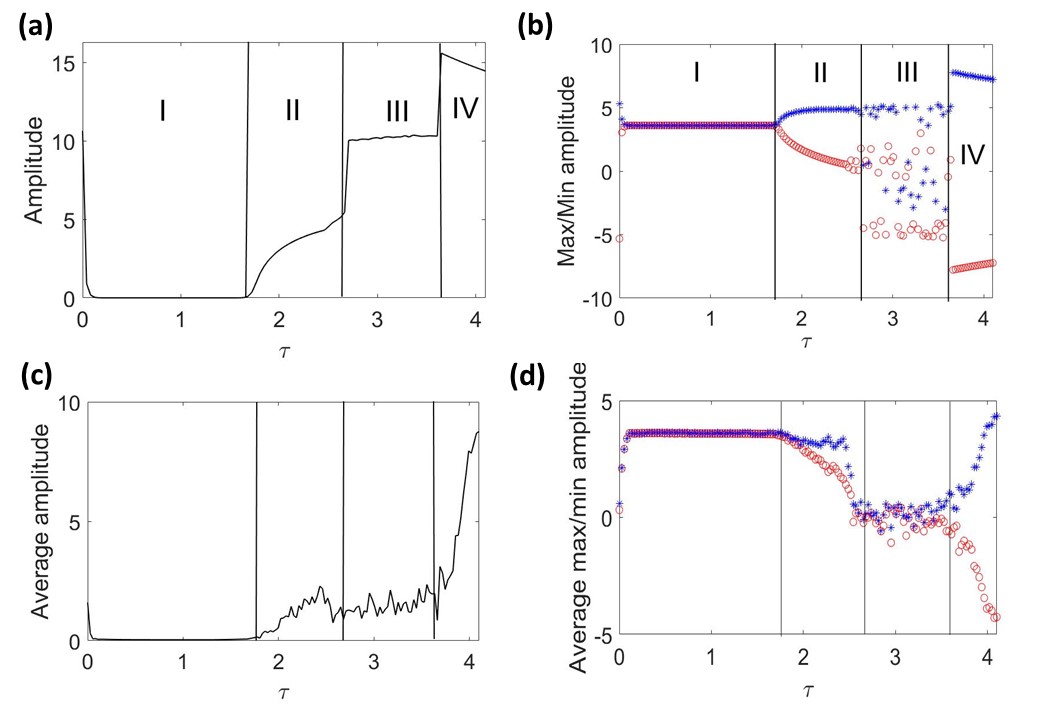}
\caption{(color online). Panels (a) and (b) depict the maximum peak-to-peak amplitude, and the maximum and minimum values of the spatial displacement as a function of the time delay for the noise-free scenario (i.e. $D_{wn}=D_{ou}=0$), respectively. The blue star points represent the maximum value of $x(t)$, whereas the red circle points correspond to its minimum values. The discussed region are indicated and separated by vertical straight lines. Similarly, panels (c) and (d) illustrate the average maximum peak-to-peak amplitude, and the average maximum and minimum values of the spatial displacement as a function of the time delay in presence of noise with $D_{wn}=D_{ou}=0.001$. Notice that the latter were obtained after taking the average over an ensemble composed of $N_{sim}=200$ trajectories. We have chosen the values of the noises as follows $C_{wn}=1$, $\tau_{ou}=1$, $C_{ou}=2/\tau_{ou}$ and $\xi^{\infty}_{ou}=0$; and the rest of parameters: $\mu=0$, $\alpha=1$, $\beta =0.1$, $\gamma=-0.3$, $F=0$, and initial conditions $x_{0}=y_{0}=1$.} \label{Fig_Tau_NNoise}
\end{figure*}

In this section, we summarize the main results concerning the dynamics of the free-noise system: the forced, time-delayed Duffing oscillator with vanishing noise. Remind that this study was extensively performed in Refs.~\cite{cantisan20201,coccolo20211}. As anticipated above, it was found that the (linear) stability of the equilibrium points $x^{\ast}_{0}$, $x^{\ast}_{1}$, and $x^*_{2}$ significantly depend of $\tau$ and $\gamma$ (when $F=0$). In particular, the curve for which $x^{\ast}_{0}$ loses stability was computed for $\tau >0$ in the overdamped case: this corresponds to a Hopf and pitchfork bifurcation for $\gamma<-1$ and $\gamma=-1$~\cite{cantisan20201}, respectively. The stability analysis was extended to the underdamped case (studied here) for $x_{1,2}^{*}$ by taking into account the dissipative effects in \cite{coccolo20211}. It turns out that the bottom of both potential wells also become unstable beyond certain values of the time delay. For our present purposes we have focused on the parameter set in which the system exhibits a rich nonlinear dynamics: in particular, we shall fix $\mu=0$, $\alpha=-1$, $\beta=0.1$ and $\gamma=-0.3$. For this choice, $x^{\ast}_{0}$ remains unstable whereas $x_{1,2}^{*}\approx \pm 3.6$  are stable for $\tau<1.76$, which resembles the well-studied bistable situation in absence of time delay \cite{rajasekar20161,jeevarathinam20111}. 

Additionally, it was shown in~\cite{cantisan20201,coccolo20211} that we can distinguish four regions in terms of the time-delay dynamics of the maximum amplitude for vanishing dissipative effects  (i.e. $\mu=0$). These regions are indicated in Fig.~\ref{Fig_Tau_NNoise}(a) which depicts the maximum peak-to-peak oscillation amplitude as a function of $\tau$ (in the free-noise system). A quick glance reveals that the oscillator eventually decays to one of the equilibrium points $x_{1,2}^{*}$ (as the maximum oscillations amplitude goes to $0$) in region I (i.e. $\tau\in(0,1.76)$), whereas its value is nonzero in regions II, III, and IV that indicates an oscillatory dynamics sustained by the time delay. In particular, it was found that the time-delayed Duffing oscillator in region III exhibits aperiodic interwell oscillations~\cite{cantisan20201,coccolo20211}. This can be manifested by means of the maximum $x_{max}$ and minimum $x_{min}$ value of the spatial displacement in each oscillation: while periodic oscillations return single well-defined values of $x_{max}$ and $x_{min}$, aperiodic oscillations give rise to an irregular distribution of points. It is important to realize that $x_{max}$ and $x_{min}$ represent half of the previous peak-to-peak amplitude, to avoid confusion we shall refer to them as either maximum or minimum amplitudes. Fig.~\ref{Fig_Tau_NNoise} (b) depicts $x_{max}$ and $x_{min}$ as a function of $\tau$, notice that the values of both magnitudes exhibit an irregular distribution when $\tau$ is in region III (i.e. $\tau\in [2.68,3.6]$). This feature is consistent with the fact that the time-delayed Duffing oscillator chaotically transits from one well to the other ~\cite{coccolo20211,cantisan20201}, we shall employ the Poincar\'e map composed of the maximum amplitudes (that is the phase space section for which $0<x$ and $\dot x \approx 0$ holds), see Fig.~\ref{Fig_PoncareMap} (g-i), to better visualize such aperiodic dynamics. This situation contrasts with region IV where the oscillator describes interwell periodic oscillations that results in a limit cycle in phase space spanning both wells. The latter can be observed from Fig.~\ref{Fig_Tau_NNoise} (b) for $\tau>3.6$: the maximum and minimum amplitudes take on regular values around $x_{1}^{*}$ and $x_{2}^{*}$, respectively. Finally, a periodic oscillatory dynamics confined in a single well arises in region II (notice that the curve of maximum and minimum amplitudes in Fig.~\ref{Fig_Tau_NNoise} (b) splits into two values around $x_{1}^{\ast}$ for $\tau\in [1.76,2.68)$). Interesting enough, it has been shown that the time-delayed Duffing oscillator can also describe a periodic interwell motion in region II by applying a driving force with identical frequency,  which is refereed to as delay-induced resonance~\cite{coccolo20211,cantisan20201}. Regarding the dissipative effects, it turns out that the above dynamics can be substantially degraded or eventually wiped out, for instance, a sufficiently large damping eventually suppresses the oscillations sustained by the time delay in region IV ~\cite{coccolo20211}. Since we are mainly interested in studying the impact of the noise upon the dynamics induced by the time delay (e.g. the interwell oscillations), it is convenient to restrict ourselves to the scenario of vanishing dissipation.

To gain some preliminary intuition about the noise effects upon the time-delayed dynamics, here we also depict the peak-to-peak amplitude, as well as the maximum and minimum values of the spatial displacement, for a small noise strength $D_{ou}=D_{wn}=0.001$ in Figs.~\ref{Fig_Tau_NNoise} (c-d). Specifically, the latter displays the mean values of these quantities after averaging over the trajectory ensemble as a function of the time delay. A direct comparison with the free-noise scenario reveals that the time-delayed dynamics is robust to sufficient weak noise effects, except in region III where the oscillation amplitudes apparently cancel (notice that in Figs.~\ref{Fig_Tau_NNoise} (c-d) the peak-to-peak amplitude as well as the maximum and minimum amplitudes seem to decay to zero for $\tau\in [2.68,3.6]$). We anticipate that this feature is a consequence of the statistic rather than some drastic change in the underlying aperiodic dynamics (for further details, see the discussion in Sec. ~\ref{SecReginoIII}). Our main concern is to study the interplay between both perturbative noises and the time delay in absence and presence of a driving force, this task is extensively carried out in the following section. In particular, a main finding is that an approximately periodic interwell motion arises in region II via stochastic resonance (see Sec.~\ref{SecReginoII} for further details). In other words, the time-delayed oscillator is no longer confined in a single well and can transit forward and backward from one well to the another driven by the combined action of the noise and the periodic force.

\section{Numerical results}
As anticipated above, in this section we perform an extensive numerical analysis of the ensemble average of the maximum peak-to-peak oscillations amplitude response and the characteristic frequency of the oscillator. The former is numerically calculated by subtracting the minimum of the times series to the maximum for each unravelling of the noisy trajectory, and then finding the mean value over all the ensemble composed of noisy $N_{sim}=100$ trajectories. The characteristic frequency is computed by using the Fourier transform of the stationary times series for each noise trajectory, by means of the Fast Fourier Transform algorithm implemented in Matlab and based on a library called FFTW. After doing this, we similarly compute the average over the aforementioned ensemble of noisy trajectories.

\begin{figure*}
\centering
\includegraphics[scale=0.28]{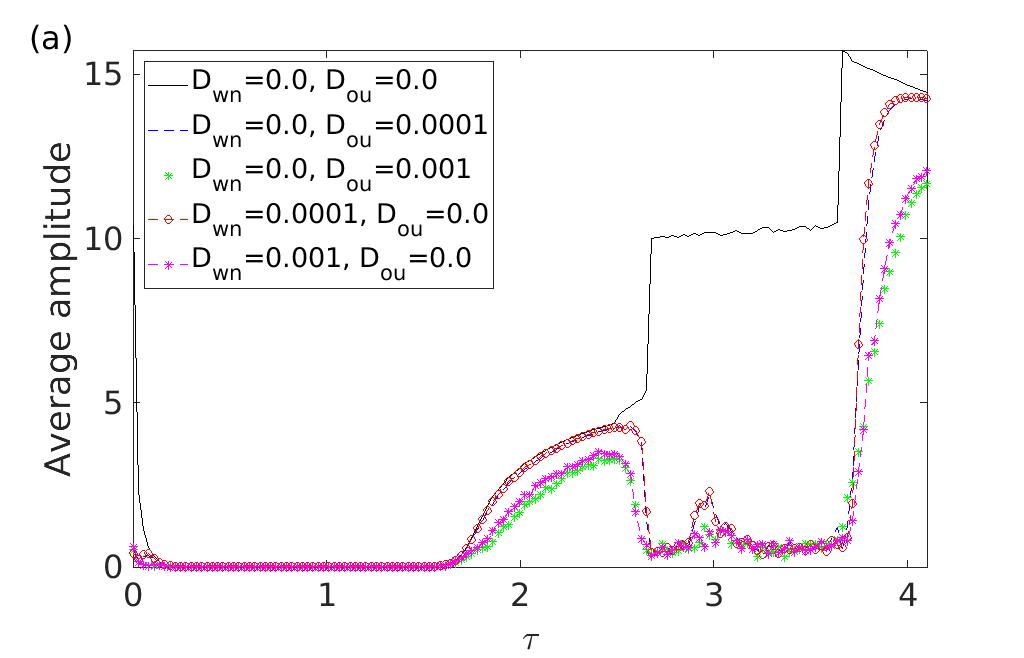}
\includegraphics[scale=0.28]{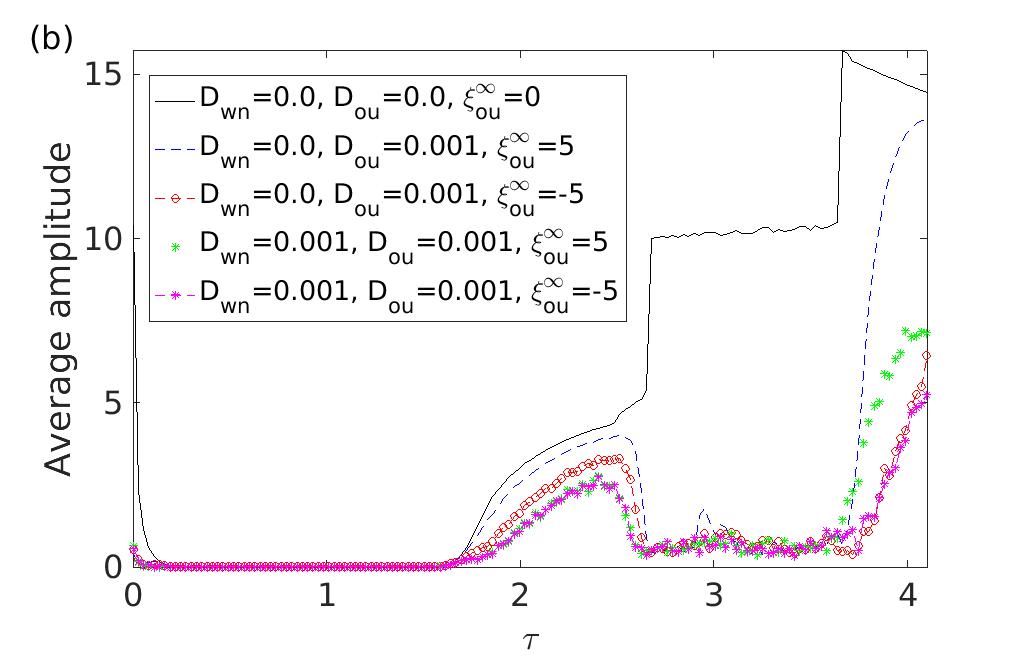}
\includegraphics[scale=0.28]{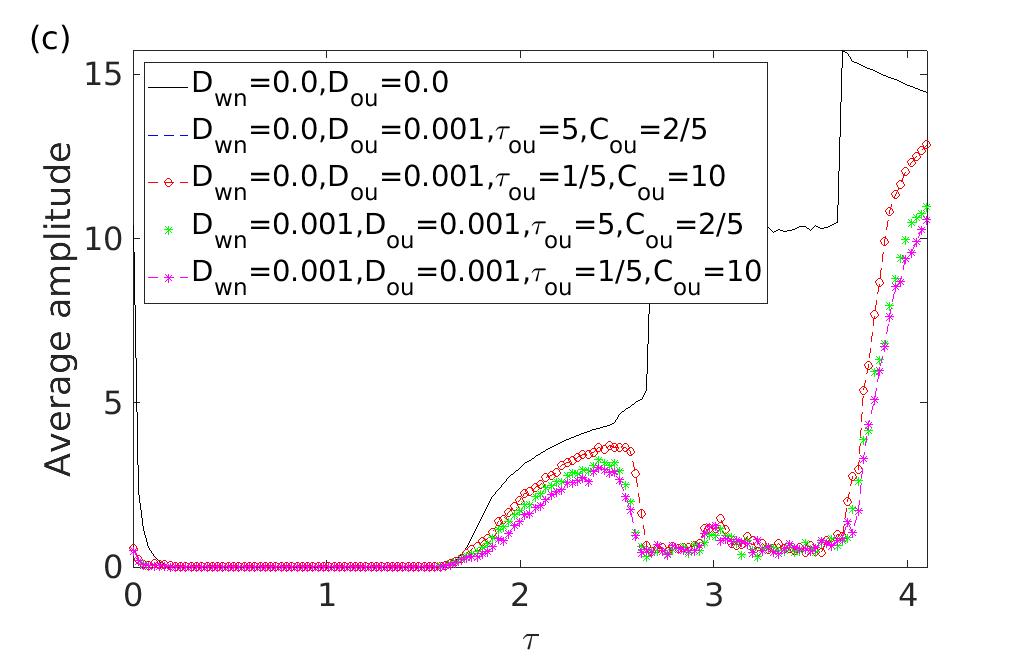}
\caption{(color online). Maximum oscillations peak-to-peak amplitude as a function of the time delay for different values of the noise parameters: Panels (a) and (b) depict the scenario with zero (i.e. $\xi_{ou}^{\infty}=0$) and non-zero time-asymptotic mean noise, respectively, for fixed values $\tau_{ou}=1$ and $C_{ou}=2$; and the panel (c) illustrates the maximum oscillations amplitude for different values of the colored-noise relaxation time $\tau_{ou}$ for vanishing time-asymptotic mean noise. We have fixed $\mu=0$, $\alpha=1$, $\beta =0.1$, $\gamma=-0.3$, and initial conditions $x_{0}=y_{0}=1$.} \label{Fig_Tau_Noise}
\end{figure*}

\subsection{Driving force \texorpdfstring{$F=0$}{F=0}}\label{SecVF}

\begin{figure*}
\centering
\includegraphics[scale=0.7]{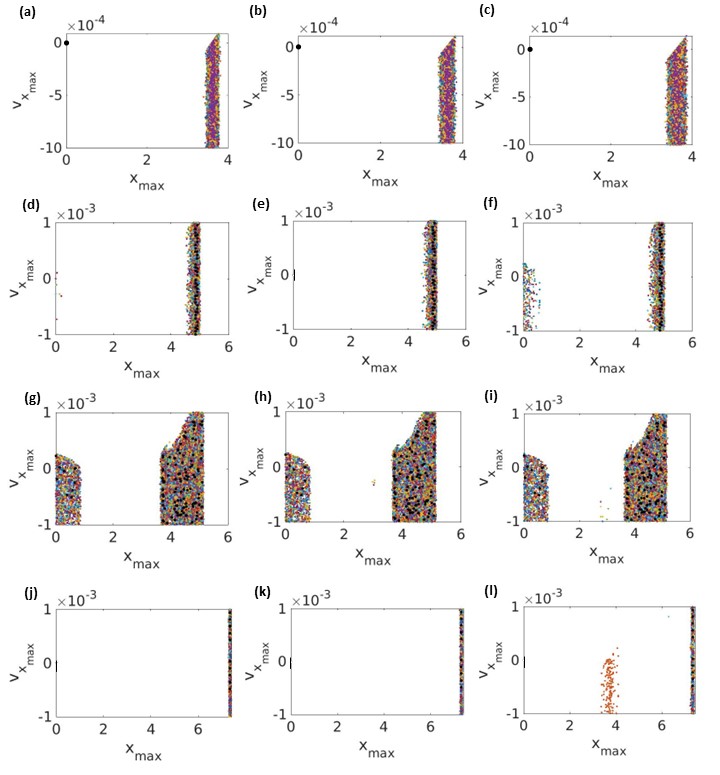}
\caption{(color online). Maximum map for three noisy scenarios and fixed values of the time delay belonging to each aforementioned region. Notice that panels in the same row have identical $\tau$: specifically, $\tau=1$, $\tau=2.5$, $\tau=3$ and $\tau=4$ in descending order; whereas panels in the same column correspond to identical noisy scenario: $D_{wn}=0.001$ and $D_{ou}=0$, $D_{wn}=0$ and $D_{ou}=0.001$, and $D_{wn}=D_{ou}=0.001$ from the left to the right. The black dots correspond to the largest displacement $x_{max}$ of each oscillation with vanishing noise, while each colored dot represents $x_{max}$ for a given unravelling of noise. We have considered an ensemble composed of $N_{sim}=200$ trajectories, and we have chosen the values of the noises as follows $C_{wn}=1$, $\tau_{ou}=1$, $C_{ou}=2/\tau_{ou}$ and $\xi^{\infty}_{ou}=0$; and the rest of parameters: $\mu=0$, $\alpha=1$, $\beta =0.1$, $\gamma=-0.3$, and initial conditions $x_{0}=y_{0}=1$.} \label{Fig_PoncareMap}
\end{figure*}

To get a useful intuition about the impact of the noisy dynamics, we start studying the maximum peak-to-peak oscillations amplitude as a function of $\tau$ for different values of the time-asymptotic noise mean $\xi_{ou}^{\infty}$, the colored noise memory time $\tau_{ou}$, and the colored noise diffusive amplitude $C_{ou}$. To make a fair comparison between the white and colored noises, we have fixed $C_{wn}=C_{ou}\tau_{ou}/2$ such that the variance of both noises is identical (see the autocorrelation function in Eqs.~(\ref{EQWN}) and (\ref{EQOUN})). 

Figures from~\ref{Fig_Tau_Noise}(a) to \ref{Fig_Tau_Noise}(c) illustrate the average values of the maximum peak-to-peak oscillations amplitude as a function of the time delay for several choices of the colored-noise parameters, which are indicated in the header of the plots.  We shall analyse the situation for small strength $D_{ou}$, $D_{wn}$ of both noises in comparison to the height of the potential barrier, i.e., $D_{ou},D_{wn}\ll \Delta U$. In all figures, it can be seen that the oscillations amplitude is barely modified for a sufficiently weak noise strength (recall that the background solid black line corresponds to the situation in absence of both noises), except in region III where it substantially decays as a direct consequence of the ensemble average, as anticipated in the previous section. The fact that the maximum peak-to-peak amplitude cancels in region III can be intuitively understood by realizing that, on one hand, the oscillator is expected to transit between both wells as occurs for zero noise values, and on the other hand, we have chosen noises such that they have no prevalence for neither of these wells (notice that $\langle \xi_{wn}(t)\rangle=0$ and $\langle \xi_{ou}(t)\rangle=0$ when $t\rightarrow \infty$), such that there is an effective cancellation in the average among the trajectories of the ensemble. This point will be further discussed in Sec.~\ref{SecReginoIII} when we study the time series of the spatial coordinate in the region III. In particular, one may also see in Fig.~\ref{Fig_Tau_Noise}(a) that the impact of both noises is similar for a given identical strength: that is, the blue (which just takes account colored noise effects) and red (which corresponds to white noise) lines as well as red and pink lines largely coincides in all regions. Additionally, Fig.~\ref{Fig_Tau_Noise}(a) reveals that the maximum oscillations amplitude in the scenarios II and IV is almost equally degraded by both noises for an identical noise strength. Indeed, we shall show in Sec.~\ref{SecReginoII} that the maximum oscillations amplitude decreases equally fast for increasing values of either the white or colored noise.

From Fig.~\ref{Fig_Tau_Noise}(b), one may appreciate that a colored noise endowed with a non-zero mean has an asymmetric effect in absence of the white noise. This feature is particularly manifested in region IV: While the oscillations amplitude takes larger values for $\xi_{ou}^{\infty}=5$ (see the blue dashed line in Fig.~\ref{Fig_Tau_Noise}(b)) in comparison with the zero-mean situation (see the green star line in Fig.~\ref{Fig_Tau_Noise}(a)), it is substantially degraded when $\xi_{ou}^{\infty}=-5$ (see the red dot-dashed line in Fig.~\ref{Fig_Tau_Noise}(b)). This result is against one could expect as the oscillator describes a limit cycle containing both wells. This asymmetric feature is substantially suppressed by the white noise, as one could appreciate from the pink star-dashed and green star lines in regions II and IV.  

Figure~\ref{Fig_Tau_Noise}(c) reveals that the OU-noise relaxation time $\tau_{ou}$ does not play a significant role in the time-asymptotic maximum oscillations amplitude either in presence or absence of the white noise: notice that the blue dashed and red dot-dashed lines completely coincide for all regions. Basically, there appears just small discrepancies in the region IV. 

 Additionally, we assess the chaotic feature of the noisy dynamics by computing the maximum maps in the three noisy scenarios: (i) $D_{wn}\neq 0$ and $D_{ou}=0$, (ii) $D_{wn}= 0$ and $D_{ou}\neq 0$, and (iii) $D_{wn}\neq 0$ and $D_{ou}\neq 0$. These are illustrated in Figs.~\ref{Fig_PoncareMap}(a-l) for a particular instance of $\tau$ embedded in each aforementioned regions (i.e. $\tau=1,2.5,3,4$). These plots consists of depicting the values of $x_{max}$ in the phase space for each realization of noise: basically, they can be understood as the points of the Poincar\'e section for which $0<x$ and $v=\dot x\approx 0$ holds. More specifically, we require that the velocity in the maximum amplitude satisfies $|v_{x_{max}}|<10^{-3}$ because we are not able to compute the strict Poincar\'e section due to numerical limitations. The latter would correspond to the set of points all lying in the X axis, instead we observe that points stack on vertical bands. In order to make a direct comparison with the free-noise results summarized in Sec.~\ref{SecPre}, we also plot the maximum amplitude returned by the noiseless situation, which are represented by the black dots.    

By paying attention to the maximum maps one may roughly distinguish two behaviours: that is, (i) the colored points are either slightly scattered around the free-noise black points, or (ii) they are broadly spread throughout the phase space. The former situation are clearly displayed by Figs.~\ref{Fig_PoncareMap}(d-f) and (j-l), where the vast majority of points are completely embedded in a narrow (vertical) band located near the equilibrium point $x_{1}^{\ast}$. These results corresponding to the regions II and IV are consistent with Fig.~\ref{Fig_Tau_NNoise}(d). A similar situation is obtained for the region I, though the colored points are not centered around the noiseless result (recall that $x(t)=0$ when $t\rightarrow\infty$ for $\tau=1$) since the noise is responsible for sustaining a oscillatory dynamics (see Sec.~\ref{SecReginoI} for further details). Accordingly, these results can be intuitively attributed to the fact that the nonlinear dynamics must be regular: the maximum amplitude remains slightly disturbed despite the oscillator trajectory is randomly changed by the noise at each time step (notice that the band width is small compared to the situation for $\tau=3$). By contrast the maximum amplitude in region III diffuses along the X axis occupying a wider band of phase space, from $x_{max}=0$ to $x_{max}\approx 0.9$ as well as from $x_{max}\approx 3.6$ to $x_{max}\approx 5.2$. This can be traced back to the fact that the underlying dynamics is chaotic, such that a small random perturbation (recall that $D_{wn},D_{ou}\ll \Delta U$) may produce a significant variation of the maximum amplitude. Furthermore, notice that this feature is appreciated in the three noisy scenarios, that is the impact of both white and colored noise upon the maximum amplitude is similar for a given identical strength, which coincides with the results observed in Figs. ~\ref{Fig_Tau_Noise}. In other words, Figs.~\ref{Fig_PoncareMap}(g-i) confirm that the chaotic dynamics intrinsic to the region III,
previously reported in Refs.~\cite{cantisan20201,coccolo20211}, also manifests in presence of either the white or colored noise. 

Special attention deserves the results for $\tau =2$ in the worst noisy scenario, see Fig.~\ref{Fig_PoncareMap}(f). Interestingly, at a closer inspection, one may observe that $x_{max}$ takes values near the origin as well, this is an indication of incoherent interwell transitions (as it similarly occurs for $\tau=3$). Concretely, these points correspond to trajectories that immediately recross the potential barrier once the oscillator has jumped to the potential well located at $x_{2}^{\ast}$ (and which eventually get back to the potential well placed in $x_{1}^{\ast}$). As previously commented, we shall show in Sec.~\ref{SecReginoII} that the noise also induces an aperiodic forward and backward hopping between potential wells in region II, that eventually turns into an approximately periodic interwell motion via stochastic resonance by applying a driving force. Finally, we should discuss the orange points located around $x_{1}^{\ast}\approx 3.6$ for the instance of region IV. Although it is not shown here, we find that such points correspond to a single trajectory which gets trapped within the well located at $x_{1}^{\ast}$. This is also a manifestation that the time-delayed dynamics may be substantially modified for sufficiently large values of noise.

\subsection{Driving force \texorpdfstring{$F\neq 0$}{F no 0}}\label{SecNVF}

Having gained some understanding of the noise influence upon the maximum peak-to-peak oscillations amplitude, we now turn the attention to the scenario in presence of a driving force.  To provide a comprehensive analysis of the impact of noises upon the time-delayed Duffing oscillator, we now present a numerical study of both the maximum peak-to-peak oscillations amplitude of the time series and the characteristic frequency in the aforementioned different scenarios of the oscillator dynamics when it is subjected to an external driving force. 

\begin{figure*}
\centering
\includegraphics[scale=0.87]{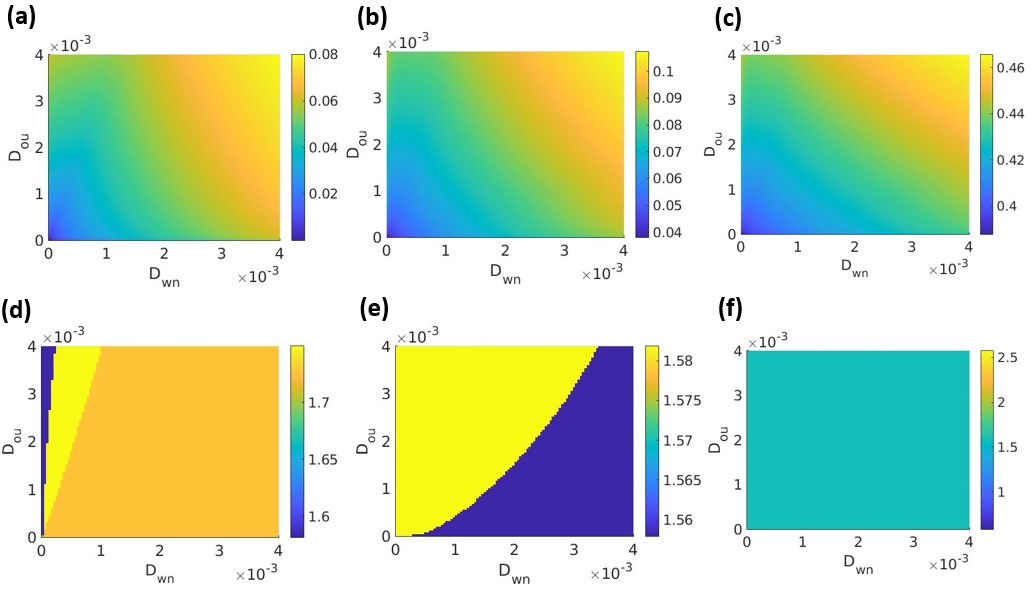}
\caption{(color online) (Upper row) Parameter set plot of the average maximum oscillations amplitude as a function of the white and colored noise strength. (Lower row) Parameter set plot of the average characteristic frequency versus the white and colored noise strength. Each column represents a fixed value of the driving force amplitude: (a) and (d) $F=0$, (b) and (e) $F=0.01$, and (c) and (f) $F=0.1$. We have chosen the values of the noises as follows $C_{wn}=1$, $\tau_{ou}=1$, $C_{ou}=2/\tau_{ou}$ and $\xi^{\infty}_{ou}=0$; and the rest of parameters: $\mu=0$, $\alpha=1$, $\beta =0.1$, $\gamma=-0.3$, $\Omega=1.571$ and initial conditions $x_{0}=y_{0}=1$.} \label{Fig_A_tau_1}
\end{figure*} 

\subsubsection{\texorpdfstring{$\tau=1$}{tau=1}}\label{SecReginoI}

We begin with $\tau=1$, which in absence of noise, corresponds to the scenario where the system is driven to one of the fixed points and the trajectories are decaying oscillations. The latter is because the time delay plays the role of a damping term in
the sense that it diminishes the oscillations caused
by the second derivative \cite{cantisan20201}.

\begin{figure*}
\centering
\includegraphics[scale=0.87]{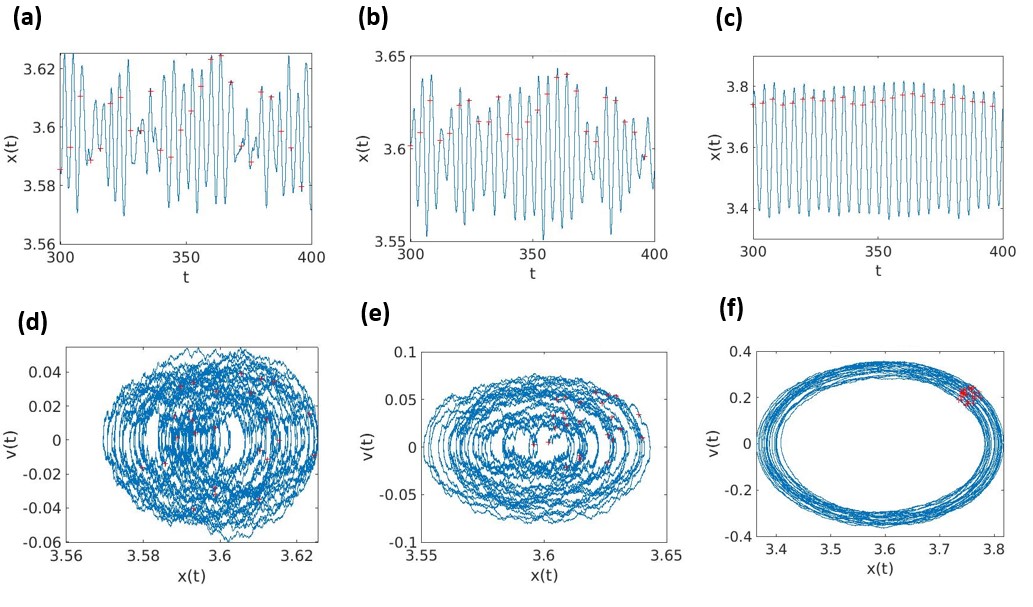}
\caption{(color online). (Upper panels) Average time series of the position as a function of the white and colored noise strength. (Lower panels) Average phase space portraits of the stochastic dynamics as a function of the white and colored noise strength. Each column represents a fixed value of the driving force amplitude: (a) and (d) $F=0$, (b) and (e) $F=0.01$, and (c) and (f) $F=0.1$. We have fixed the strength of the white and colored noise as $D_{wn}=0.004$ and $D_{ou}=0.004$; and the values of the noises as follows $C_{wn}=1$, $\tau_{ou}=1$, $C_{ou}=2/\tau_{ou}$ and $\xi^{\infty}_{ou}=0$. Similarly, we have chosen $\mu=0$, $\alpha=1$, $\beta =0.1$, $\gamma=-0.3$, $\Omega=1.571$ and initial conditions $x_{0}=y_{0}=1$.} \label{Fig_X_tau_1}
\end{figure*}

Figures~(\ref{Fig_A_tau_1})(a-c) illustrate the maximum peak-to-peak oscillations amplitude and Figs.~(\ref{Fig_A_tau_1})(d-f) depict the characteristic frequency as functions of the strength of the white and colored noises for three values of the driving force: for strong driving ($F=0.1$), for middle driving ($F=0.01$) and in absence of driving ($F=0$). It can be seen from the upper panels that the oscillations amplitude gets larger values by increasing the strength of both noises independently by the driving force. This can be intuitively understood by recalling that noises constantly inject energy to the oscillator which in turn produce an increment of the oscillations amplitude. In other words, it arises a competition between the time delay and noise effects in the region I: while the former acts as a damping mechanism, the presence of the latter to some extent sustains the oscillatory dynamics with the growth of the noise strength. Upon further inspection, one may also realize that both noises are equally responsible for such increase. As somehow expected, the driving force also contribute to the increment of the oscillations amplitude, which can be clearly seen in Figs.~\ref{Fig_X_tau_1}(a-c). The latter depicts the average position and phase space for a particular value of both noises. One may further appreciate that the oscillator movement in the phase space still manifests an underlying regular aperiodic dynamics though it becomes erratic owing to noise effects. Nonetheless, this noisy feature is significantly diminished as the driving force strength increases.  Interestingly, Fig.~\ref{Fig_X_tau_1}(c) reveals that a sort of noisy limit cycle eventually emerges due to the competition between noise and forcing effects: that is, a sufficient large driving force eventually restores the regular and periodic motion on average. One could be tentative to refer to Fig.~\ref{Fig_X_tau_1}(f) as attractor, however this represents an average value of the noisy trajectory rather than the set of states (in the phase space) of the real trajectory for a particular unraveling of the noise. Additionally, the particle path circles randomly around this limiting average trajectory in the asymptotic time, which implies that it is not closed in contrast to ordinary limit cycles.

Regarding the characteristic frequency, Figs.~\ref{Fig_A_tau_1}(d-f) also reveal that this ultimately converges to the driving force frequency $\Omega=1.571$ as the force amplitude increases (recall that, in absence of noise, the dynamics becomes periodic due to the driving force). The latter is unveiled by the red crosses in Figs.~\ref{Fig_X_tau_1}(d-f), which depicts the average position for given values of the strength of the driving force and noises. More specifically, the red crosses represents time periodic points with period $T_{\Omega}=\frac{2\pi}{\Omega}$, and thus, the fact that they are horizontally aligned in Fig.~\ref{Fig_X_tau_1}(f) manifests that the oscillator characteristic frequency coincides with $\Omega$. One may thus conclude that the driving force substantially suppresses the noise effects.

\subsubsection{\texorpdfstring{$\tau=2.5$}{tau=2.5}}\label{SecReginoII}

Now we turn the attention to the second region, which is characterized for sustaining an oscillatory dynamics that is confined in one of the wells.

\begin{figure*}
\centering
\includegraphics[scale=0.87]{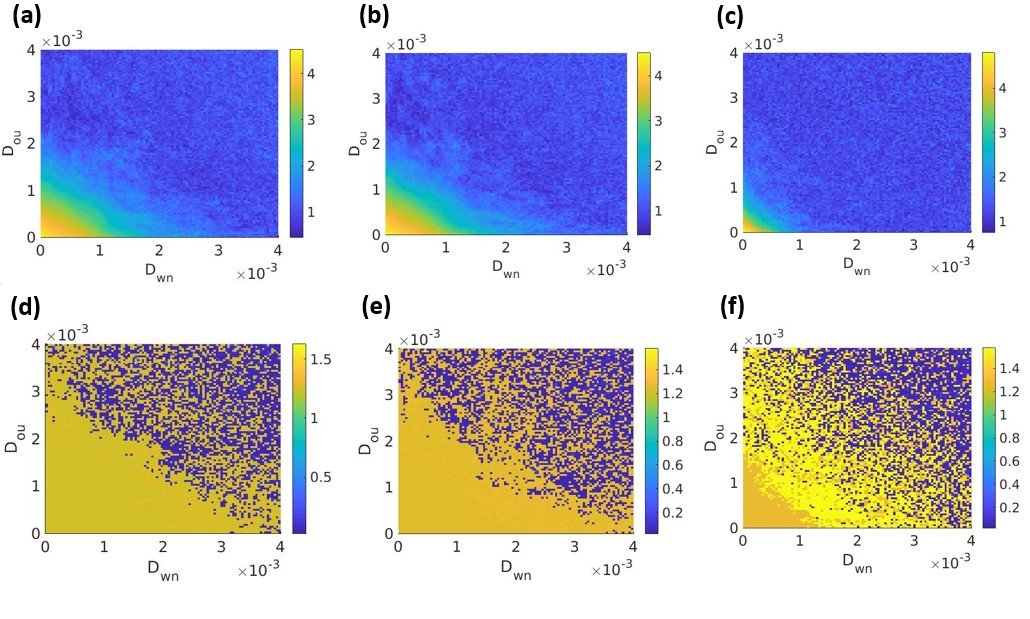}
\caption{(color online) Parameter set plot of the average maximum oscillations amplitude (upper row) and the average characteristic frequency (lower row) as functions of the strength of the white and colored noises for given values of the driving force amplitude: (a) and (d) $F=0$, (b) and (e) $F=0.01$, (c) and (f) $F=0.1$. We have chosen the values of the noises as follows $C_{wn}=1$, $\tau_{ou}=1$, $C_{ou}=2/\tau_{ou}$ and $\xi^{\infty}_{ou}=0$; and the rest of parameters: $\mu=0$, $\alpha=1$, $\beta =0.1$, $\gamma=-0.3$, $\Omega=1.571$ and initial conditions $x_{0}=y_{0}=1$.} \label{Fig_A_tau_25}
\end{figure*}

Figure~\ref{Fig_A_tau_25} depicts the average oscillations amplitude (upper row) as well as the average characteristic frequency (lower row) when we set the time delay $\tau=2.5$. Unlike the previous case, the oscillations amplitude now decays as the noise strength grows, which was anticipated by Fig.~\ref{Fig_Tau_Noise}(a). Remarkably, this degrading effect pronounces for larger values of the driving force: notice that the yellow region (representing the highest amplitude) in Figs.~\ref{Fig_A_tau_25}(a-c) shrinks simultaneously for increasing strength of either white or colored noises. Further, from Figs.~\ref{Fig_A_tau_25}(d-f), one can appreciate that the characteristic frequency decreases as a consequence of the noisy dynamics. Nonetheless, the driving force tends to sustain the periodic dynamics endowed with certain frequency close to forcing $\Omega$ since the yellow domain spreads abroad the parameter noise space.  

\begin{figure*}
\centering
\includegraphics[scale=0.87]{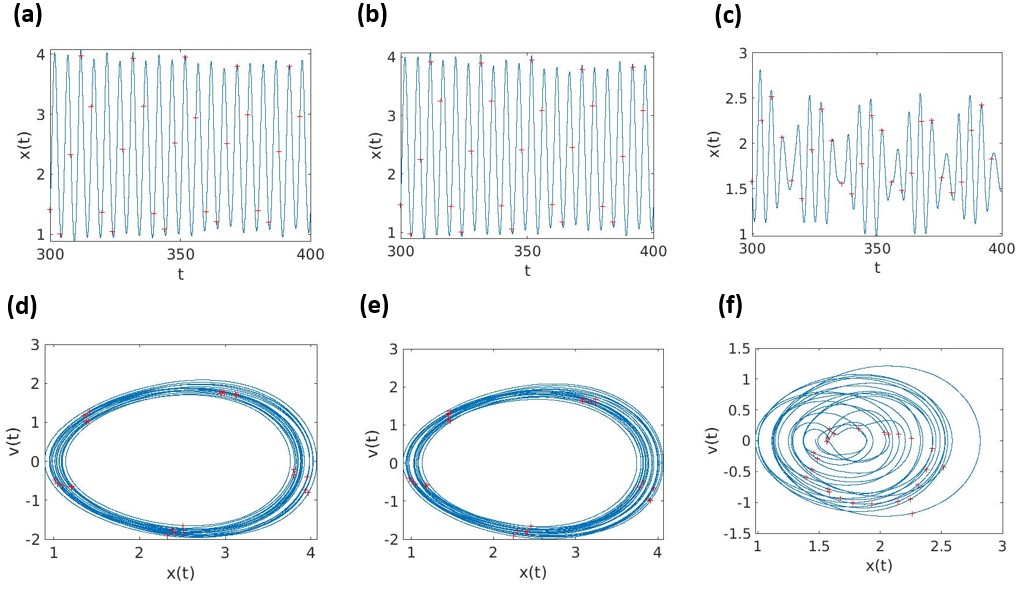}
\caption{(color online). (Upper panels) Average time series of the position as a function of the white and colored noise strength. (Lower panels) Average phase space representation as a function of the white and colored noise strength.  Each column represents a fixed value of the driving force amplitude: (a) and (d) $F=0$, (b) and (e) $F=0.01$, and (c) and (f) $F=0.1$. We have fixed the strength of the white and colored noise as $D_{wn}=0.0005$ and $D_{ou}=0.0005$;  and the values of the noises as follows $C_{wn}=1$, $\tau_{ou}=1$, $C_{ou}=2/\tau_{ou}$ and $\xi^{\infty}_{ou}=0$. Additionally, we have chosen $\mu=0$, $\alpha=1$, $\beta =0.1$, $\gamma=-0.3$, $\Omega=1.571$ and initial conditions $x_{0}=y_{0}=1$.} \label{Fig_X_tau_25}
\end{figure*}

By paying attention to the average time series (upper panels) and the phase space dynamics (lower panels) depicted in Fig.~\ref{Fig_X_tau_25}, one may appreciate that the combination of the driving force together with the noise effects give rise to an incipient irregular dynamics enable to destroy the typical limit cycle in region II \cite{cantisan20201}: we observe from the phase portraits depicted by Figs.~\ref{Fig_X_tau_25}(e-f) that the edge of the limit cycle gets wider as the dynamics becomes erratic owing to the noise effects, that are significantly powered after switching on the driving force. This result is also reflected in the characteristic frequency above, which is eventually destroyed by increasing the strength of either the white or the colored noise. Hence, the noise combined with the driving force leads to ultimately destroy the limit cycle in the region II. Interestingly, examining closely Fig.~\ref{Fig_X_tau_25}(c), this also provides useful insight about the spatial coordinate dynamics when subject to a sufficiently strong driving force: this clearly resemblances an aperiodic dynamics with varying amplitude. 

\begin{figure*}
\centering
\includegraphics[scale=0.5]{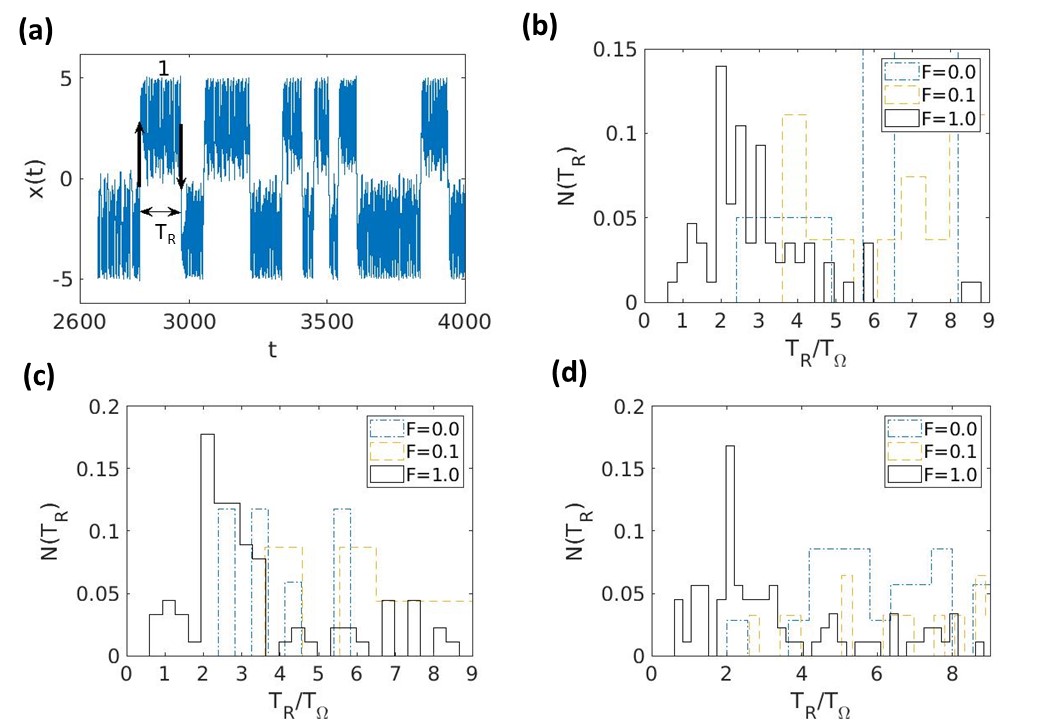}
\caption{(color online). Panel (a) depicts the time series of the spatial coordinate for a given realization of the white noise with $D_{wn}=0.01$ (i.e. $D_{ou}=0$). The forward and backward jumps over the potential barrier $\Delta U$ is indicated by the black arrows, and $T_{R}$ represent the time interval between two consecutive hoppings. Panels (b), (c) and (d) illustrate the residence-time distribution as a normalized histogram for the three noisy scenarios: (b) $D_{wn}=0.01$ and $D_{ou}=0$, (c) $D_{wn}=0$ and $D_{ou}=0.01$, and (d) $D_{wn}=D_{ou}=0.01$. Notice that the latter were obtained for an ensemble composed of $N_{sim}=500$ trajectories and sufficient large times $T_{\infty}=500 T_{\Omega}$. We have fixed $C_{wn}=1$, $\tau_{ou}=1$, $C_{ou}=2/\tau_{ou}$ and $\xi^{\infty}_{ou}=0$; and the rest of parameters $\mu=0$, $\alpha=1$, $\beta =0.1$, $\gamma=-0.3$, $\Omega=1.571$ and initial conditions $x_{0}=y_{0}=1$.} \label{Fig_Res_Time}
\end{figure*}

Differently from the other regions, we also found here that for sufficiently large noise strength (e.g. $D_{ou},D_{wn}> 0.002\Delta U$), we observe that the noise makes possible for the oscillator to pass over the potential barrier even in absence of driving force, which is better illustrated by Fig.~\ref{Fig_Res_Time}(a). Unlike previous figures, the latter depicts the time series of $x(t)$ for a given realization of the white noise (i.e. it just displays the values returned by a single instance of the ensemble trajectory rather than its average). Clearly, one may appreciate that the oscillator initially transits from the  potential well located below $x_{2}^{\ast}\approx -3.6$ to the other one, and gets back after certain time denoted by $T_{R}$ (which is refereed to as the residence time \cite{gammaitoni19981}). This represents an entire interwell transition (i.e. a forward and backward hopping between both potential wells), and it can repeat a number $N_{H}$ of times for a given unraveling of the noise in any of the three noisy scenarios previously discussed. As somehow expected, we find that the latter increases with the strength of the driving force, this is illustrated in Fig.~\ref{Fig_NTrac}(d-f) in App.~\ref{App2}. Here, we show the $N_H$ in function of the ratio between the forcing amplitude and the potential barrier. Furthermore, we find that $N_{H}$ also depends of the noise correlation time $\tau_{ou}$: notice that, by paying attention to Figs.~\ref{Fig_NTrac}(e) and (f), such growth significantly diminishes as $\tau_{ou}$ gets too small or large compared with $\tau$. Additionally, we also study the percentage of noisy trajectories $N_{T}$ (within the ensemble) that manifests at least a single complete transition, which is depicted in Figs. ~\ref{Fig_NTrac}(a-c) as a function of $F$. This quantity displays a similar behavior to $N_{H}$, for instance, $N_{T}$ substantially grows in presence of both $\xi_{wn}$ and $\xi_{ou}$ for comparatively small values of the noise strength, which means that the probability to observe an oscillatory interwell motion rises with the driving force. In other words, this feature suggests that the time interval between the forward and backward hoppings, that is $T_{R}$, may get synchronized with the periodic forcing time $T_{\Omega}$ by means of stochastic resonance \cite{gammaitoni19981}. In order to delve into this question we have also computed the residence-time distribution, denoted by $N(T_{R})$, in the three noisy scenarios and for values of the forcing amplitude $F=0,0.1,1$. This is displayed in Figs. ~\ref{Fig_Res_Time}(b-d) as a normalized histogram with finite size bins.

We observe that a sharp peak around twice the forcing period arises as the forcing amplitude takes on values sufficiently large compared to the potential barrier (i.e. $F=1\approx 0.24 \Delta U$), which can be viewed as a fingerprint of synchronization, i.e. the preferred residence-time is apparently $T_{R}\approx 2T_{\Omega}$. By contrast, we can further appreciate that the residence-time distribution is significantly wide for sufficiently small or vanishing driving force. This can be intuitively understood by realizing that most of the noise-induced transitions must be incoherent (i.e. they are erratic interwell transitions) so that any value of $T_{R}$ has approximately equal likelihood. Let us recall that $N(T_{R})$ has been previously employed to characterize the stochastic resonance in (non-time-delayed) bistable systems in presence of the white noise \cite{gammaitoni19891,choi19981}. In particular, it is well-known that the condition of synchronization in those systems corresponds to $T_{R}=T_{\Omega}/2$ (and odd multiples) \cite{gammaitoni19981}. This result can be traced back to the fact that the potential is most extremely tilted to the right or the left at times $(n-\frac{1}{2})T_{\Omega}$, which in turn promotes the interwell transition. Compared with our problem, it is difficult to figure out a simple picture due to the many time scales that involves our problem (e.g. the time delay $\tau$, the noise correlation time $\tau_{ou}$, etc). Nonetheless, one may, at least, expect that the synchronization condition may change with respect to the non-delayed situation as a consequence of the additional time scale indirectly introduced by $\tau$ in the potential (\ref{PtFunc}).

\subsubsection{\texorpdfstring{$\tau=3$}{tau=3}}\label{SecReginoIII}

In this section, we examine the situation in which the oscillator can transit from one fixed point to the other and the dynamics is chaotic \cite{cantisan20201}. Here, the time delay makes the oscillator jump between the potential wells describing aperiodic trajectories.

\begin{figure*}
\centering
\includegraphics[scale=0.87]{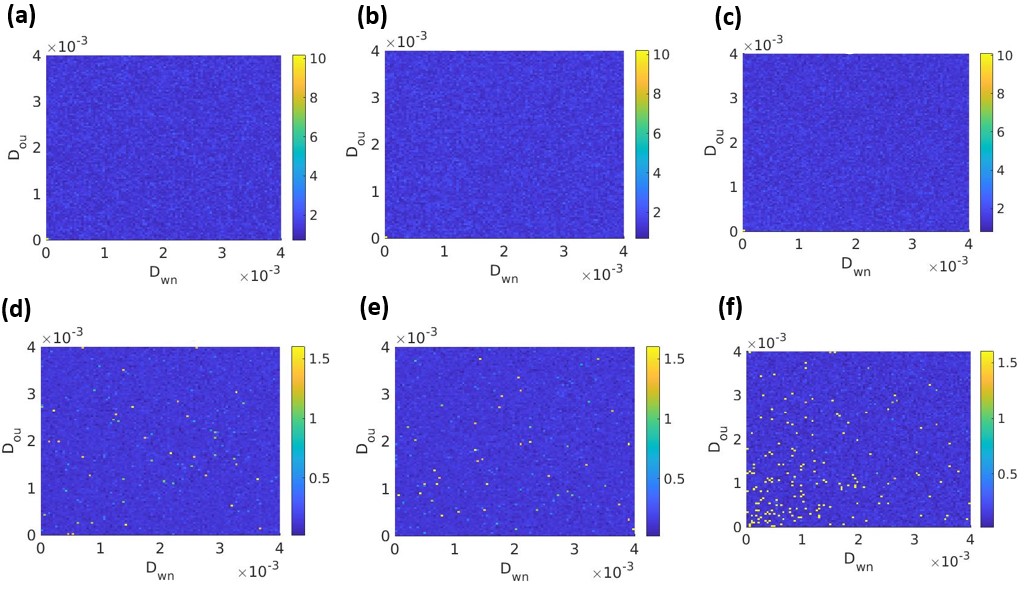}
\caption{(color online) Parameter set plot of the average maximum oscillations amplitude (upper row) and the average characteristic frequency (lower row) as functions of the strength of the white and colored noises for given values of the driving force amplitude: (a) and (d) $F=0$, (b) and (e) $F=0.01$, (c) and (f) $F=0.1$. We have chosen the values of the noises as follows $C_{wn}=1$, $\tau_{ou}=1$, $C_{ou}=2/\tau_{ou}$ and $\xi^{\infty}_{ou}=0$; and the rest of parameters: $\mu=0$, $\alpha=1$, $\beta =0.1$, $\gamma=-0.3$, $\Omega=1.571$ and initial conditions $x_{0}=y_{0}=1$.} \label{Fig_A_tau_3}
\end{figure*}

\begin{figure*}
\centering
\includegraphics[scale=0.87]{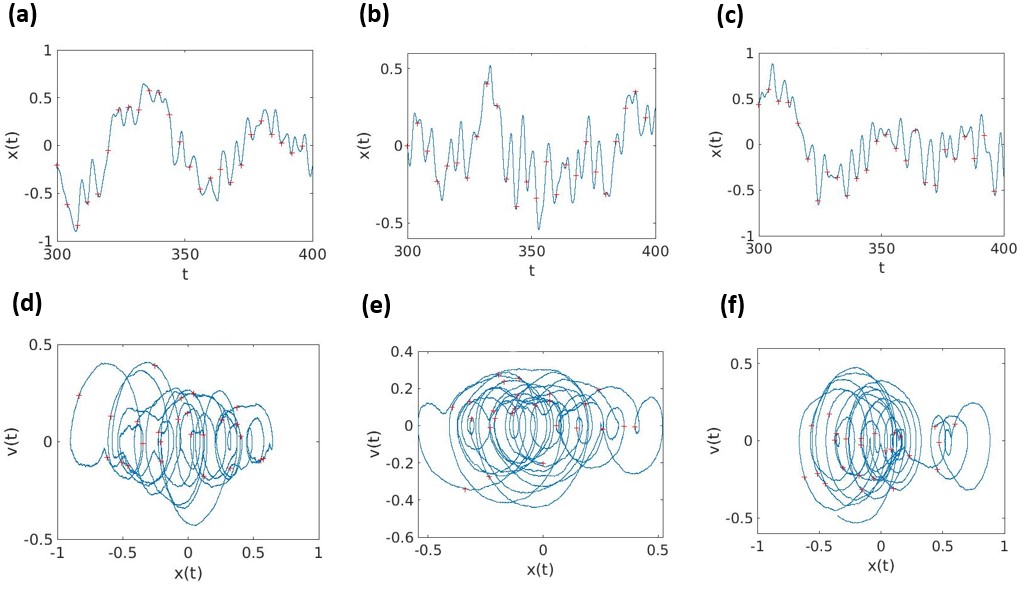}
\caption{(color online). (Upper panels) Average time series of the position as a function of the white and colored noise strength. (Lower panels) Average phase space representation as a function of the white and colored noise strength.  Each column represents a fixed value of the driving force amplitude: (a) and (d) $F=0$, (b) and (e) $F=0.01$, and (c) and (f) $F=0.1$. We have fixed the strength of the white and colored noise as $D_{wn}=0.0005$ and $D_{ou}=0.0005$; and the values of the the noises as follows $C_{wn}=1$, $\tau_{ou}=1$, $C_{ou}=2/\tau_{ou}$ and $\xi^{\infty}_{ou}=0$. Similarly, we have chosen $\mu=0$, $\alpha=1$, $\beta =0.1$, $\gamma=-0.3$, $\Omega=1.571$ and initial conditions $x_{0}=y_{0}=1$.} \label{Fig_X_tau_3}
\end{figure*}

As was anticipated in Sec.~\ref{SecVF}, Figs.~\ref{Fig_A_tau_3}(a-c) show that the average maximum oscillations amplitude decays to zero for arbitrary nonzero values of the noise strength in either the presence or absence of the driving force. To understand this result is convenient to pay attention to the ensemble average of the time series and the phase space dynamics illustrated in Fig.~\ref{Fig_X_tau_3}. Although it is not shown here, a single unraveling of the noisy dynamics reveals that the oscillator describes an aperiodic interwell motion (i.e. it truly transits between both wells). Nonetheless, Figs.~\ref{Fig_X_tau_3}(a-c) show that the oscillator is apparently restricted to move around the origin, which certainly leads to a misunderstanding. This is rooted on the fact that the white and colored noises do not discern between both wells (recall that $\langle \xi_{wn}(t)\rangle=0$ and $\langle \xi_{ou}(t)\rangle=0$ when $t\rightarrow \infty$) such that the vast majority of the noisy trajectories compensate each other after doing the ensemble average: that is, the oscillator turns to be symmetrically located with respect to the origin most of the time for any two unravellings of the noise. In other words, certain spatial displacement $-x(t)$ and $x(t)$ have the same chance for a given $t$ in the trajectory ensemble. This misleading feature is also reflected in the phase space by Figs.~\ref{Fig_X_tau_3}(d-f), which shows that the ensemble average dynamics takes places around the origin rather than embedding both potential wells. In short, though Figs.~\ref{Fig_A_tau_3}(a-c) resemble an oscillation death of the dynamics, this is just due to the fact that there is no prevalence for any of the potential wells due to our choice of the noise parameters.  An observation should be made here: this feature could be observed in the amplitude plots Figs. ~\ref{Fig_A_tau_25}(a-c) for the instance in region II as well, where the stochastic resonance is the responsible for the interwell transitions, so that one could expect that the average of the peak-to-peak amplitude would decrease for a enough large noise strength. 

Notice that the maximum peak-to-peak oscillations amplitude for zero noise values and driving force is consistent with previous works~\cite{cantisan20201} (i.e., it is equal to the double of the length between potential wells), though it is difficult to appreciate at first sight from Fig.~\ref{Fig_A_tau_3}(a). In conclusion, the noise effects make the averaging dynamics completely irregular for $\tau =3$ such that the oscillations get scattered along the parameter set.

\subsubsection{\texorpdfstring{$\tau=4$}{tau=4}}

Finally, we study the case in which the oscillator is no longer confined to either of the wells and its trajectory corresponds in a limit cycle in the phase space that contains both wells.

\begin{figure*}
\centering
\includegraphics[scale=0.87]{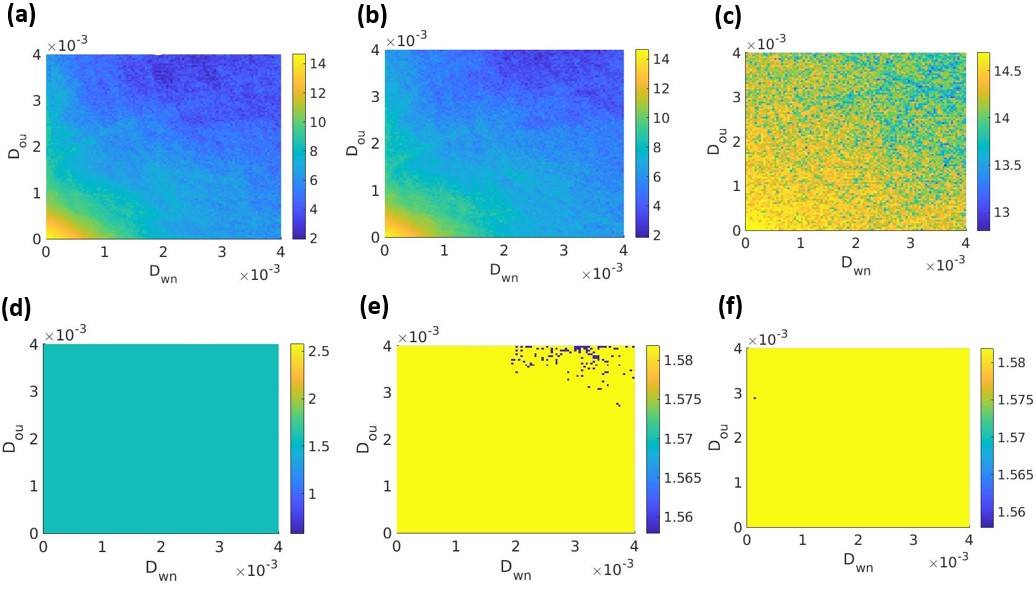}
\caption{(color online) Parameter set plot of the average maximum oscillations amplitude (upper row) and the average characteristic frequency (lower row) as functions of the strength of the white and colored noises for given values of the driving force amplitude: (a) and (d) $F=0$, (b) and (e) $F=0.01$, (c) and (f) $F=0.1$. We have chosen the values of the noises as follows $C_{wn}=1$, $\tau_{ou}=1$, $C_{ou}=2/\tau_{ou}$ and $\xi^{\infty}_{ou}=0$; and the rest of parameters: $\mu=0$, $\alpha=1$, $\beta =0.1$, $\gamma=-0.3$, $\Omega=1.571$ and initial conditions $x_{0}=y_{0}=1$.}\label{Fig_A_tau_4}
\end{figure*}

\begin{figure*}
\centering
\includegraphics[scale=0.87]{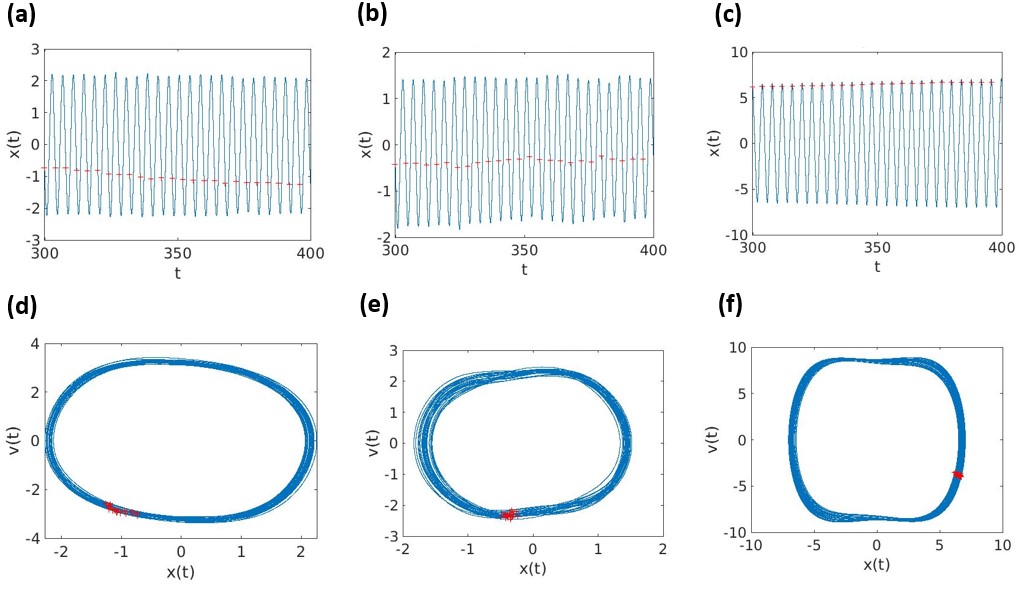}
\caption{(color online).(Upper panels) Average time series of the position as a function of the white and colored noise strength. (Lower panels) Average phase space representation as a function of the white and colored noise strength.  Each column represents a fixed value of the driving force amplitude: (a) and (d) $F=0$, (b) and (e) $F=0.01$, and (c) and (f) $F=0.1$. We have fixed the strength of the white and colored noise as $D_{wn}=0.004$ and $D_{ou}=0.004$; and the values of the noises as follows $C_{wn}=1$, $\tau_{ou}=1$, $C_{ou}=2/\tau_{ou}$ and $\xi^{\infty}_{ou}=0$. Additionally, we have chosen $\mu=0$, $\alpha=1$, $\beta =0.1$, $\gamma=-0.3$, $\Omega=1.571$ and initial conditions $x_{0}=y_{0}=1$.} \label{Fig_X_tau_4}
\end{figure*}

From Figs.~\ref{Fig_A_tau_4}(a-c), one can observe that the maximum oscillations amplitude of such oscillatory dynamics is significantly diminished by a growing strength of the white and colored noises as anticipated in Sec.~\ref{SecVF}. Although this situation seems to resemble the one of region II, a sufficiently strong forcing is now able to reestablish the maximum peak-to-peak oscillations amplitude: in contrast to Fig.~\ref{Fig_A_tau_25}(c), focusing on Fig.~\ref{Fig_A_tau_4}(c), the yellow region spreads over all the parameter set when $F=0.1$. This feature is also reflected by the time series of the spatial coordinate, see Figs.~\ref{Fig_X_tau_4}(a-c), where it can be observed that the amplitude of the oscillatory dynamics eventually grows with the driving force amplitude. While the maximum oscillations amplitude is rather vulnerable to the noise effects, the characteristic frequency proves to be fairly resilient against them. This is clear from Figs.~\ref{Fig_A_tau_4}(d-f), which manifests that the characteristic frequency remains unchanged despite the increment of the noise strength. In particular, the characteristic frequency converges to $\Omega$ as indicated by the red crosses plotted in the times series of the spatial coordinate (see Figs.~\ref{Fig_X_tau_4}(d-f). As similarly occurred in region I (see Fig.~\ref{Fig_A_tau_1}(c)), the nonlinear oscillator is ultimately tuned to the frequency $\Omega$ of the driving force.

More interestingly, by paying attention to the phase space dynamics depicted in Figs.~\ref{Fig_X_tau_4}(d-f), we find out that the shape and the amplitude of the limit cycle in the present scenario (with $\tau=4$) are affected by noise independently of the forcing amplitude (i.e. its border just gets wider). This is also in contrast with the case with $\tau=2$, where a growing driving force contributes to eventually destroy the limit cycle.

\section{Outlook and conclusion}

We have extensively analyzed the maximum peak-to-peak oscillations amplitude and the characteristic frequency in the stationary state for a time-delayed Duffing oscillator affected by both white and active OU noises. In the first part, by assuming a zero driving force, we show that both noises have a similar degrading effect upon the maximum peak-to-peak oscillations amplitude independently of the time delay and for a given identical strength. We also observe that the OU-noise relaxation time does not play an important role in the time-asymptotic maximum oscillations amplitude either in presence or absence of the white noise. Then, we have switched on the driving force and find an intricate interplay among noise, forcing and time delay. In fact, we show that the influence of both noises upon the dynamics changes from one region to the other: for instance, the combination of noise and driving force in region II eventually destroys the characteristic limit cycle; whilst a regular and periodic dynamics emerges in region I by increasing values of the external forcing. Similarly, a sufficiently strong driving force is able to restore a deformed limit cycle in region IV, as well as the characteristic frequency turns to be rather resilient against noise effects. Interestingly, we find that, for sufficiently large values of the noise and the forcing amplitude, an approximately periodic interwell motion arises in region II by means of stochastic resonance: specifically, the interwell time scale and the forcing period get synchronized to promote an interwell transition in the three noisy scenarios studied here.

Active particles have proved to be useful in the description of a great diversity of phenomena appearing in biological and physical systems where time delay effects can be present. Our study apparently suggests that the impact of the active noise in the time-delayed nonlinear dynamics does not substantially distinguishes from that owing to the white noise. Nonetheless, it would be necessary to further investigate in this line to unveil the role of the active motion in the phenomena beside time-delayed nonlinear dynamics. In particular, it is appealing to study how other phenomena besides stochastic resonance are influenced by the time delay in the presence of active noise. 

\section{Acknowledgment}

This work has been supported by the Spanish State Research Agency (AEI) and the European Regional Development Fund (ERDF, EU) under Project No.~PID2019-105554GB-I00.

\appendix

\counterwithin{figure}{section}
\section{Convergence test for the trajectory number}\label{App1}
\begin{figure*}
\centering
\includegraphics[scale=0.5]{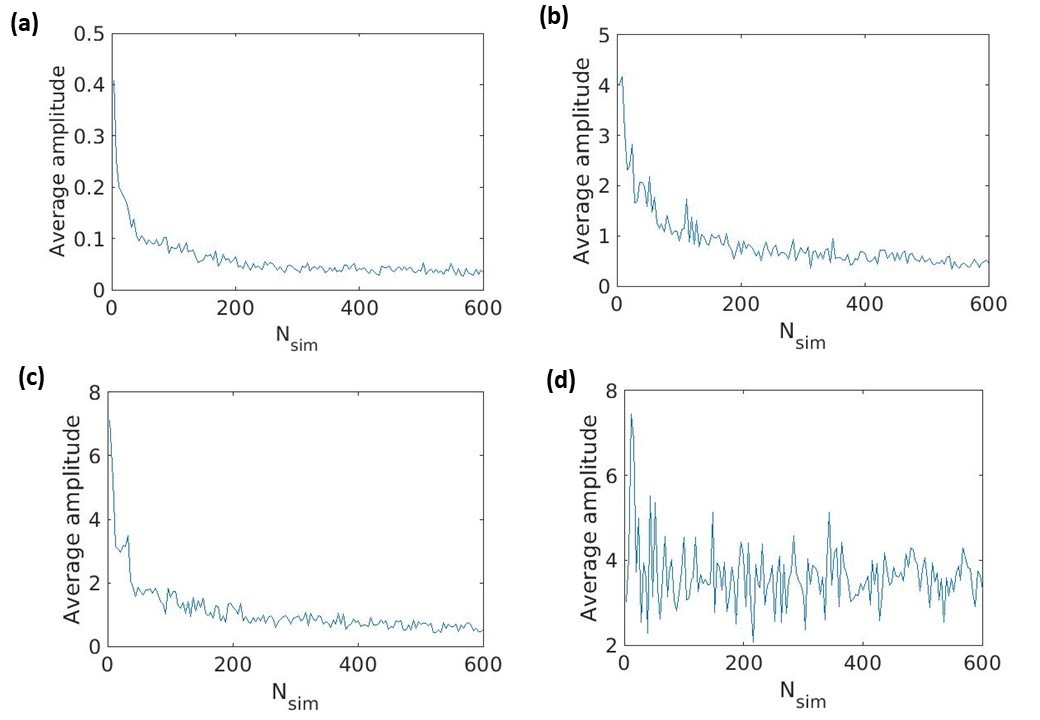}
\caption{(color online). Test of convergence of the maximum peak-to-peak oscillations amplitude in terms of the number $N_{sim}$ of the trajectory ensemble for fixed values of the noise strength in each scenario: (a), (b), (c) and (d) correspond to $\tau=1$, $\tau=2.5$, $\tau=3$, $\tau=4$; respectively. We have fixed the values of the noises as follows $D_{wn}=D_{ou}=0.004$, $C_{wn}=1$, $\tau_{ou}=1$, $C_{ou}=2/\tau_{ou}$ and $\xi^{\infty}_{ou}=0$. Similarly, we have chosen $\mu=0$, $\alpha=1$, $\beta =0.1$, $\gamma=-0.3$, $F=0$ and initial conditions $x_{0}=y_{0}=1$.} \label{Fig_Tau_Noise_Conv}
\end{figure*}

In this section, we provide several figures that assess the validity of our approximation about the number of trajectories introduced in Sec.~\ref{SecMMT}: that is $N_{sim}=100$. In particular, Figs.~\ref{Fig_Tau_Noise_Conv}(a-d) depict the maximum peak-to-peak oscillations amplitude as a function of the trajectory number composing the ensemble average in the different regions studied above. One may observe that the value of the maximum oscillations amplitude barely changes beyond the point $N_{sim}=100$ in all the scenarios analyzed: more specifically, it fluctuates around certain asymptotic number with a small oscillations amplitude. We also notice that these figures have been obtained for the largest values of the noise strength studied in the present paper, so we may expect that such fluctuations would be diminished for lower values of the noise. Although it is not shown here, we have obtained similar results for different choices of the colored noise parameters. Since the calculation of the ensemble average becomes computationally time consuming as $N_{sim}$ increases, the number of ensemble trajectories $N_{sim}=100$ returns a reasonably accurate  estimation of the maximum peak-to-peak oscillations amplitude for our present purposes.

\section{\texorpdfstring{$N_{T}$}{NT} and \texorpdfstring{$N_{H}$}{NH} in region II}\label{App2}
\begin{figure*}
\centering
\includegraphics[scale=0.7]{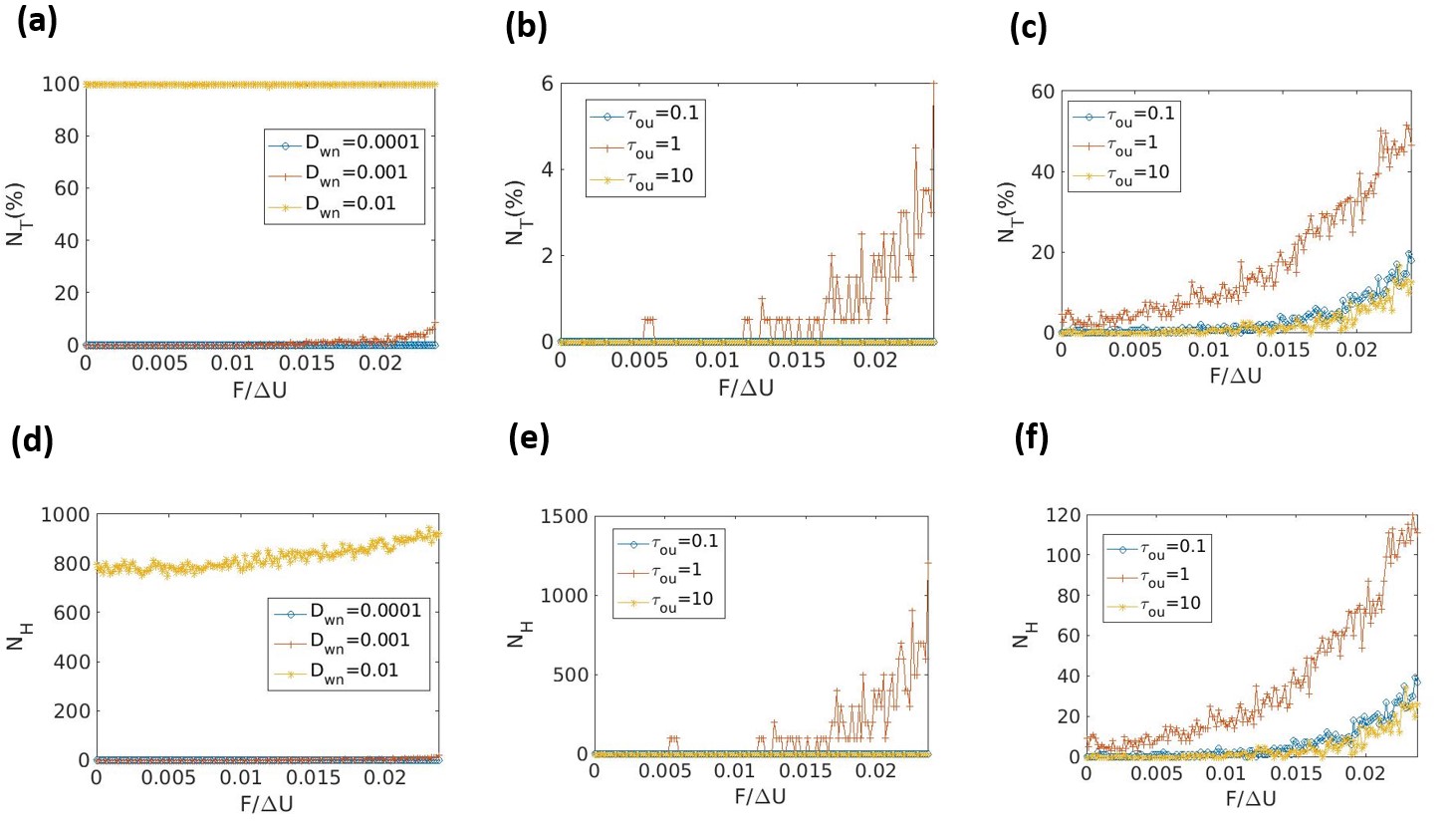}
\caption{(color online). (Upper panels) Percentage of trajectories manifesting entire interwell transitions (at least, a single forward-backward hopping) as a function of the forcing amplitude in the three noisy scenarios: (a) $D_{wn}=0.001$ and $D_{ou}=0$, (b) $D_{wn}=0$ and $D_{ou}=0.001$; and (c) $D_{wn}=D_{ou}=0.001$. (Lower panels) Number of complete transitions as a function of the forcing amplitude in the three noisy scenarios: (d) $D_{wn}=0.001$ and $D_{ou}=0$, (e) $D_{wn}=0$ and $D_{ou}=0.001$; and (f) $D_{wn}=D_{ou}=0.001$. We have fixed the strength of the white and colored noise as indicated in the title of the plots; and $C_{wn}=1$, $C_{ou}=2/\tau_{ou}$ and $\xi^{\infty}_{ou}=0$; and the rest of parameters $\mu=0$, $\alpha=1$, $\beta =0.1$, $\gamma=-0.3$, $\Omega=1.571$ and initial conditions $x_{0}=y_{0}=1$.} \label{Fig_NTrac}
\end{figure*}

Focusing on the instance in region II (i.e. $\tau=2.5$), we show the results about the percentage of trajectories within the ensemble that exhibit, at least, a forward-backward transition between both potential wells (see Figs. \ref{Fig_NTrac}(a-c)), as well as the number of complete transitions as a function of the forcing amplitude (see Figs.~\ref{Fig_NTrac}(d-f)). These plots were computed for an ensemble composed of $N_{sim}=500$ trajectories. Interestingly, we observe that the white noise induces entire interwell transitions in any ensemble trajectory for $D_{wn}=0.01$, while sufficiently large or small noise correlation times $\tau_{ou}$ diminish the number of complete transitions. Additionally, we have found that such interwell motion may get synchronized with the periodic force, as discussed in Sec. \ref{SecReginoII}.

\bibliographystyle{apsrev4-1}
\bibliography{references}
\end{document}